\documentclass[reprint,superscriptaddress,amsmath,amssymb,aip,apl,twocolumn]{revtex4-2}
\usepackage{graphicx}% Include figure files
\usepackage{dcolumn}% Align table columns on decimal point
\usepackage{bm}% bold math
\usepackage[hidelinks]{hyperref}% add hypertext capabilities
\usepackage[mathlines]{lineno}%  numbering of text and display math
\begin{document}
%-------------------------------------------------------------------------------
\title{High-performance Kerr quantum battery}
%-------------------------------------------------------------------------------
\author{Muhammad~Shoufie Ukhtary}\email{m.shoufie.ukhtary@brin.go.id}
\affiliation{Department of Physics and Astronomy, University of Exeter, EX4 4QL, United Kingdom}
%\affiliation{Department of Physics, Faculty of Mathematics and Natural Sciences, Universitas Indonesia, Depok 16424, Indonesia}
\affiliation{Research Center for Quantum Physics, National Research and Innovation Agency (BRIN), South Tangerang 15314, Indonesia}
%-------------------------------------------------------------------------------
\author{Ahmad~R.~T.~Nugraha}\email{ahmad.ridwan.tresna.nugraha@brin.go.id}
\affiliation{Research Center for Quantum Physics, National Research and Innovation Agency (BRIN), South Tangerang 15314, Indonesia}
\affiliation{Research Collaboration Center for Quantum Technology 2.0, Bandung 40132, Indonesia}
\affiliation{Department of Engineering Physics, Telkom University, Bandung 40257, Indonesia}
%-------------------------------------------------------------------------------
\author{Adam B. Cahaya}
\affiliation{Research Center for Quantum Physics, National Research and Innovation Agency (BRIN), South Tangerang 15314, Indonesia}
\affiliation{Department of Physics, Faculty of Mathematics and Natural Sciences, Universitas Indonesia, Depok 16424, Indonesia}
%-------------------------------------------------------------------------------
\author{Andrivo Rusydi}
\affiliation{Department of Physics, National University of Singapore, Singapore 117551, Singapore}
%-------------------------------------------------------------------------------
\author{Muhammad~Aziz Majidi}\email{aziz.majidi@sci.ui.ac.id}
\affiliation{Department of Physics, Faculty of Mathematics and Natural Sciences, Universitas Indonesia, Depok 16424, Indonesia}
%-------------------------------------------------------------------------------
\begin{abstract}
We propose and investigate the performance of a hybrid quantum battery, the so-called Kerr quantum battery, which consists of two interacting quantum oscillators, i.e., the charger is a harmonic oscillator and the battery is an anharmonic oscillator involving the Kerr nonlinearity.  Such a setup creates nonuniform spacing between energy levels of the quantum oscillator that increases with the energy level.  We find that the Kerr quantum battery can store more energy than the qubit battery and reaches maximum stored energy faster than the harmonic oscillator battery.  In particular, the average charging power of the Kerr quantum battery is larger than the qubit battery.  Furthermore, most of the stored energy in the Kerr quantum battery can be extracted for work.  All of the properties of the Kerr quantum battery are controlled by the strength of nonlinearity, in which the enhancement of the nonlinearity transforms the battery from a harmonic oscillator to a qubit.
\end{abstract}

\date{\today}
\maketitle
% \section{Introduction}
% \label{sec:int}

Traditionally, one can define a battery as a physical system that stores energy supplied by an external source and then makes it available to be used by other devices.  We can also characterize the battery's performance by several indicators such as (1) the amount of energy stored in the battery, (2) the energy that can be used for work, and (3) the charging time.   Although battery research seems to already evolve into a mature technological product, rapid advances in nanoscience and nanotechnology have recently triggered a new interest in the so-called ``quantum battery", which is an energy-storing system that exploits quantum effects, such as quantum entanglement and correlations, to obtain better performances than the conventional (classical) batteries~\cite{Alicki2013,Acin2013,Binder2015,Campaioli2017,Le2018,Henao2018}.  The quantum battery, in principle, relies on quantum resources to obtain extraordinary performance in energy storage and transfer~\cite{Campaioli2017, Le2018, Henao2018}.  There have been various proposals to realize the quantum battery, among which several works by Polini's group~\cite{Andolina2018, Andolina2019, Farina2019} are particularly outstanding.  Rosa's group has also pointed out the quantum advantage for several models of quantum battery~\cite{rossini2020quantum,shaghaghi2022micromasers,rodriguez2023ai,rosa2020ultra} and the fundamental limit of the maximum charging power of quantum battery~\cite{gyhm2022quantum}.  In this regard, let us consider an interface between a quantum battery and its external energy supply or the so-called ``quantum charger", which is also a quantum system.  We may expect quantum correlations between the battery and the charger to speed up the battery's charging process. 

In practical applications, where we only have access to the battery, such quantum effects, unfortunately, can also reduce the amount of energy transformed into extractable work.  To understand whether or not the quantum battery exhibits ``quantum advantage" over the classical battery, it is thus important to generalize the quantum battery and charger's model by exposing the whole system to the environment in an open-system setup~\cite{Farina2019}, which one can formulate in terms of master equations.  In such a picture, the energy is dynamically injected into the system (specifically, the charger) by an external force from the environment.  With the presence of the environment and for different implementations of the charger and battery systems, one can explicitly compute the total energy transferred to the battery and also the fraction of it, defined as ergotropy~\cite{allahverdyan04-max}, that turns out to be useful in terms of extractable work.  Farina \emph{et al.} in 2019 have proposed the combination of harmonic oscillators and two-level systems (qubits) in the quantum battery system~\cite{Farina2019}. In particular, using the harmonic oscillator as a model of both the charger and the battery provides more stored energy.  Meanwhile, using a hybrid system comprising a harmonic oscillator as the charger and a qubit as the battery gives the fastest charging.  However, the maximum stored energy in the battery becomes limited by the energy scale of the qubit. Hence, achieving a quantum battery that is capable of storing more energy with faster charging time is desirable.

% \section{Model and methods}
% \label{sec:model-methods}

\begin{figure}[t]
\centering\includegraphics[clip,width=7cm]{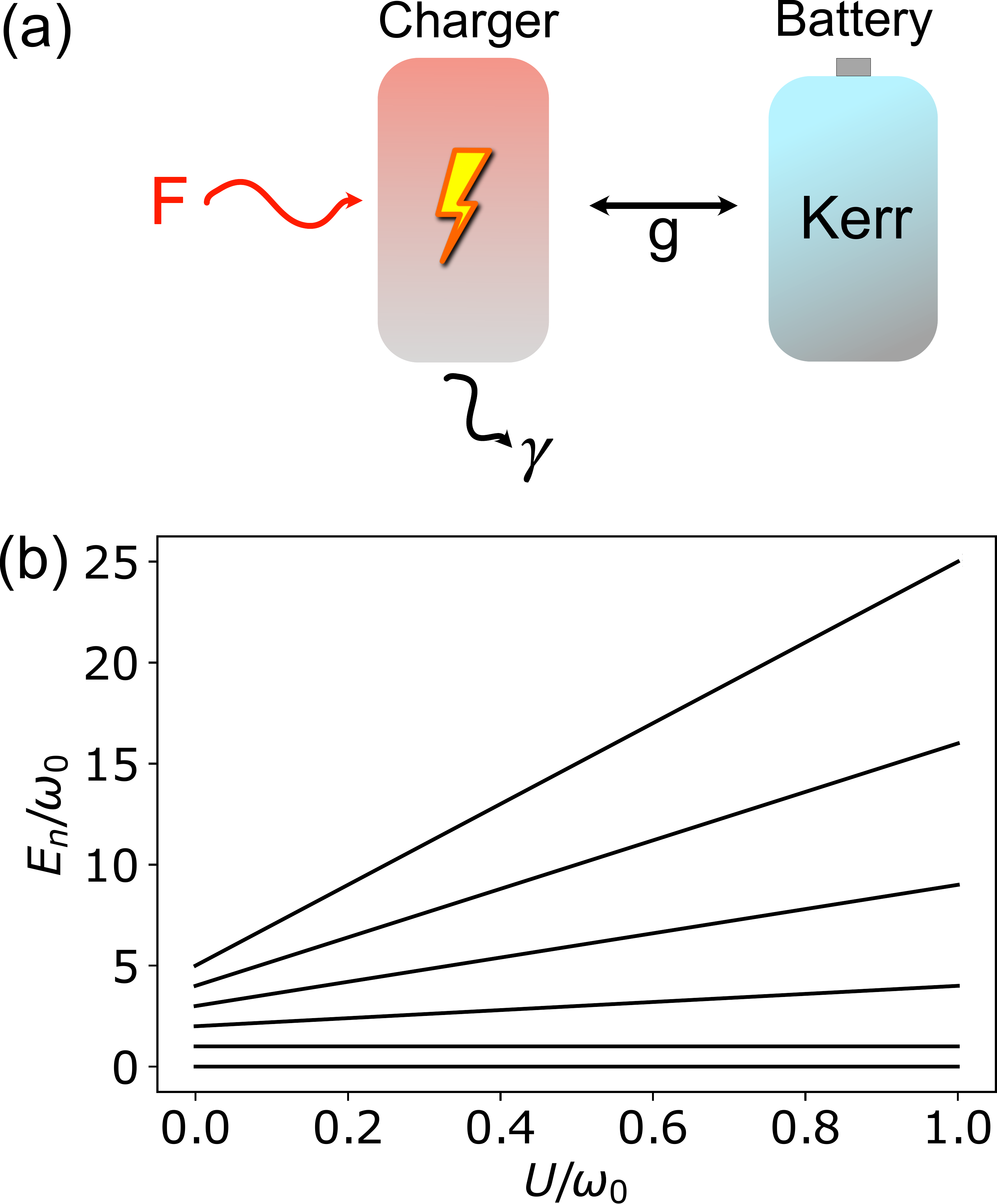}
\caption{\label{fig:bat} Proposal of ``Kerr quantum battery". (a) The Kerr quantum battery system consists of two interacting harmonic oscillators labeled as the charger and the battery, in which we include the effect of Kerr nonlinearity.  The charger is driven by the external field with a strength of $F$ and it interacts with the environment through the coupling constant $\gamma$, while $g$ represents the interaction strength between the charger and the battery.  (b) The energy levels of an oscillator with the Kerr nonlinearity.}
\end{figure}

To overcome the above-mentioned problem, in this Letter, we investigate the performance of a hybrid quantum battery that takes the benefit of faster charging time from the qubit but is also able to store more energy.  Our model is based on two interacting quantum oscillators labeled the charger and the battery as shown in Fig.~\ref{fig:bat}(a). The charger interacts with the environment allowing dissipation.  To charge the battery, the charger is driven by a coherent driving force.  The excited energy in the charger is transferred to the battery through the interaction between them.  In contrast to the charger, the battery is not driven and is well-isolated from the environment.  Thus, the task of the battery is to store the energy from the charger that later can be used for work. Here, we consider the anharmonicity in the oscillator for the battery by including the Kerr nonlinearity \cite{imamoglu1997strongly}.  Such a setup is what we call a ``Kerr (quantum) battery".  On the other hand when there is no Kerr nonlinearity, we will refer to the setup as the ordinary ``harmonic oscillator battery".  

Due to the Kerr nonlinearity, which is characterized by the $U$ parameter, the spacing between two levels of the oscillator is no longer uniform and it increases with increasing levels as shown in Fig. \ref{fig:bat}(b). For a relatively large $U$ value, the harmonic oscillator effectively becomes a two-level system or a qubit battery. For the intermediate $U$, depending on the driving strength and dissipation, the effective number of levels can be more than two. Therefore, at the intermediate $U$, we expect more stored energy in the Kerr battery than the qubit battery with yet faster charging time than the harmonic oscillator battery. In particular, Kerr nonlinearity gives additional control of the battery's properties for a fixed driving strength and dissipation. We will show that the Kerr nonlinearity transforms the properties of the battery from the harmonic oscillator to qubit features.  Note that the Kerr nonlinearity can be realized in a cavity system filled with a nonlinear medium, in which a weak-control laser determines the strength of nonlinearity $U$~\cite{imamoglu1997strongly,zhu2010large}. 

The Hamiltonian of the Kerr battery in a frame rotating with the external drive during the charging process is expressed as follows:
\begin{align}
    H =& \Delta(a^\dag a + b^\dag b)+g(ab^\dag+a^\dag b) +F(a^\dag +a)\notag\\
        &+Ub^\dag b^\dag bb,\label{eq:H}
\end{align}
where the detuning frequency is defined as $\Delta=\omega_0-\omega_D$. We consider identical oscillators for the charger and the battery with a frequency of $\omega_0$.  The frequency of external driving force is denoted by $\omega_D$. When $\Delta=0$, the driving is resonant with the system. $a(a^\dag)$ and $b(b^\dag)$ are the bosonic annihilation (creation) operators for the charger and the battery, respectively.  The second term in the Hamiltonian of Eq.~\eqref{eq:H} is the interaction between the charger and the battery with an interaction strength $g$.  Here, we employ the rotating wave approximation for the interaction term by neglecting terms associated with rapid oscillation in time $(a^\dag b^\dag +ab)$~\cite{Farina2019,farina2021dissipative,yuan2021unconventional}. The third term gives the coherent driving term with a driving strength $F$~\cite{Zhang2021,walls2008quantum,gerry2005introductory,scully_zubairy_1997}. The last term corresponds to the Kerr nonlinearity for the battery~\cite{imamoglu1997strongly,Zhang2021,zhang2021driven}.  Note that setting $U=0$ corresponds to the harmonic oscillator battery.  Therefore, the Hamiltonian in Eq.~\eqref{eq:H} is more general than that used by Farina \emph{et al.}~\cite{Farina2019} due to the inclusion of detuning and Kerr nonlinearity terms.

The interaction of the system with the environment is represented effectively by the Lindbladian term in the master equation \cite{manzano2020}.  The master equation of our system is expressed as follows:
\begin{align}
    \partial_t\rho(t)=&-i[H,\rho(t)] \notag\\
    &+\frac{\gamma}{2}(2a\rho(t)a^\dag-a^\dag a\rho(t)-\rho(t)a^\dag a)\label{eq:me},
\end{align}
where $\rho(t)$ is the density operator of the system and $\gamma$ is the dissipation constant representing the energy dissipation from the charger to the environment. The charger is in contact with a thermal bath as an environment that induces transfer of population between them~\cite{Farina2019,manzano2020}.  We consider a low-temperature setup of the bath that allows us to neglect the population transfer from the environment to the charger or thermal excitation. We also focus on the coherent driving for the excitation of the charger, since the thermal excitation does not contribute to extractable energy from the battery~\cite{Farina2019}. By solving the master equation, the expectation value of the operator can be calculated as $\langle O\rangle(t)=\textrm{Tr}[O\rho(t)]$. 

\begin{figure}
\centering\includegraphics[clip,width=7.5cm]{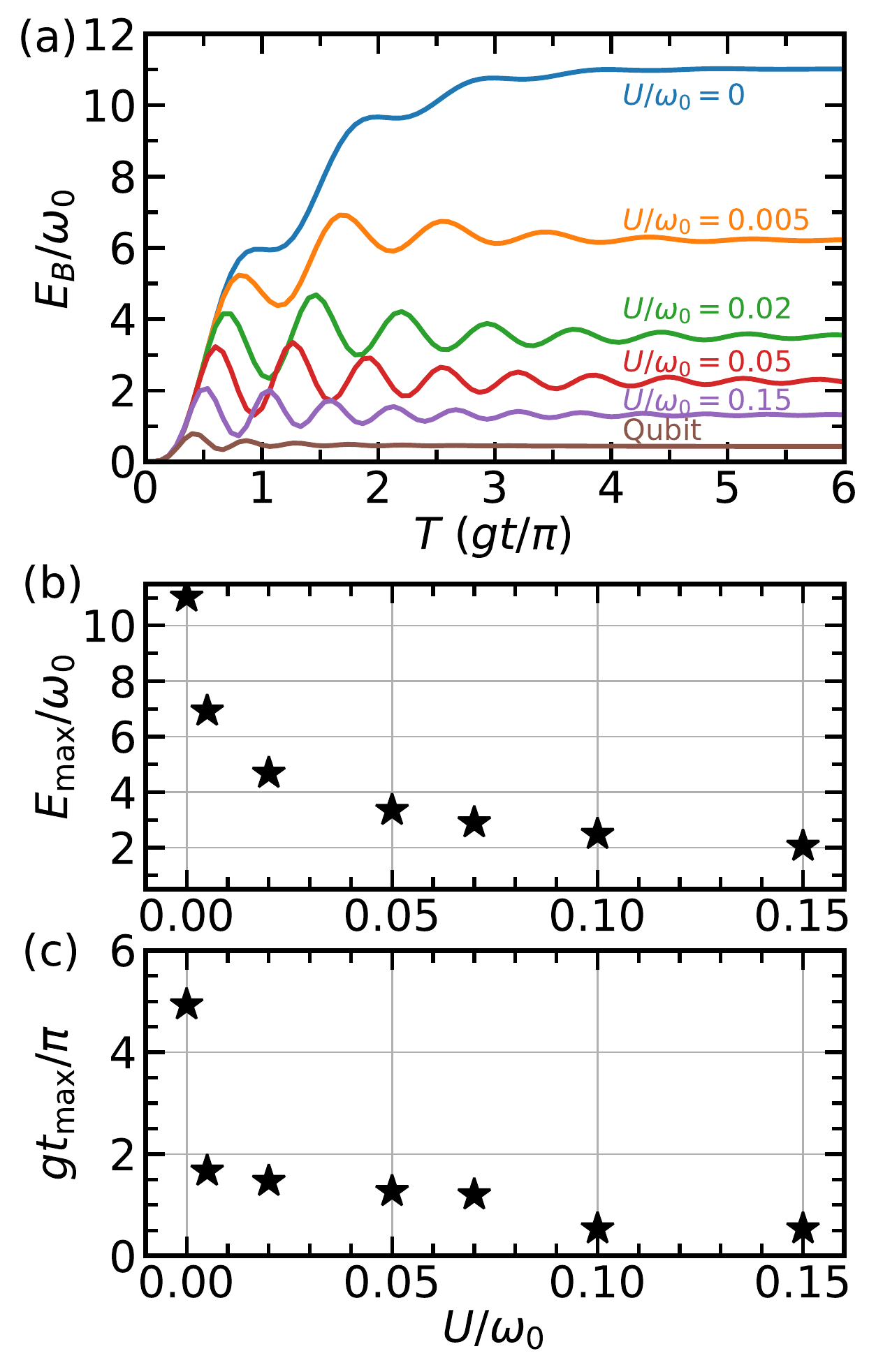}
\caption{\label{fig:ew} Some performance indicators of the Kerr quantum battery (a) $E_B/\omega_0$ as a function of charging time $T\equiv gt/\pi$ for several $U/\omega_0$. (b) The maximum $E_B/\omega_0$ for several $U/\omega_0$. (c) The optimal charging time for several $U/\omega_0$. In all cases, we set $\Delta=0.2\omega_0$, $\gamma=0.3\omega_0$, $g=0.2\omega_0$, and $F=0.5\omega_0$.}
\end{figure}

\begin{figure}
\begin{center}
\includegraphics[clip,width=7.5cm]{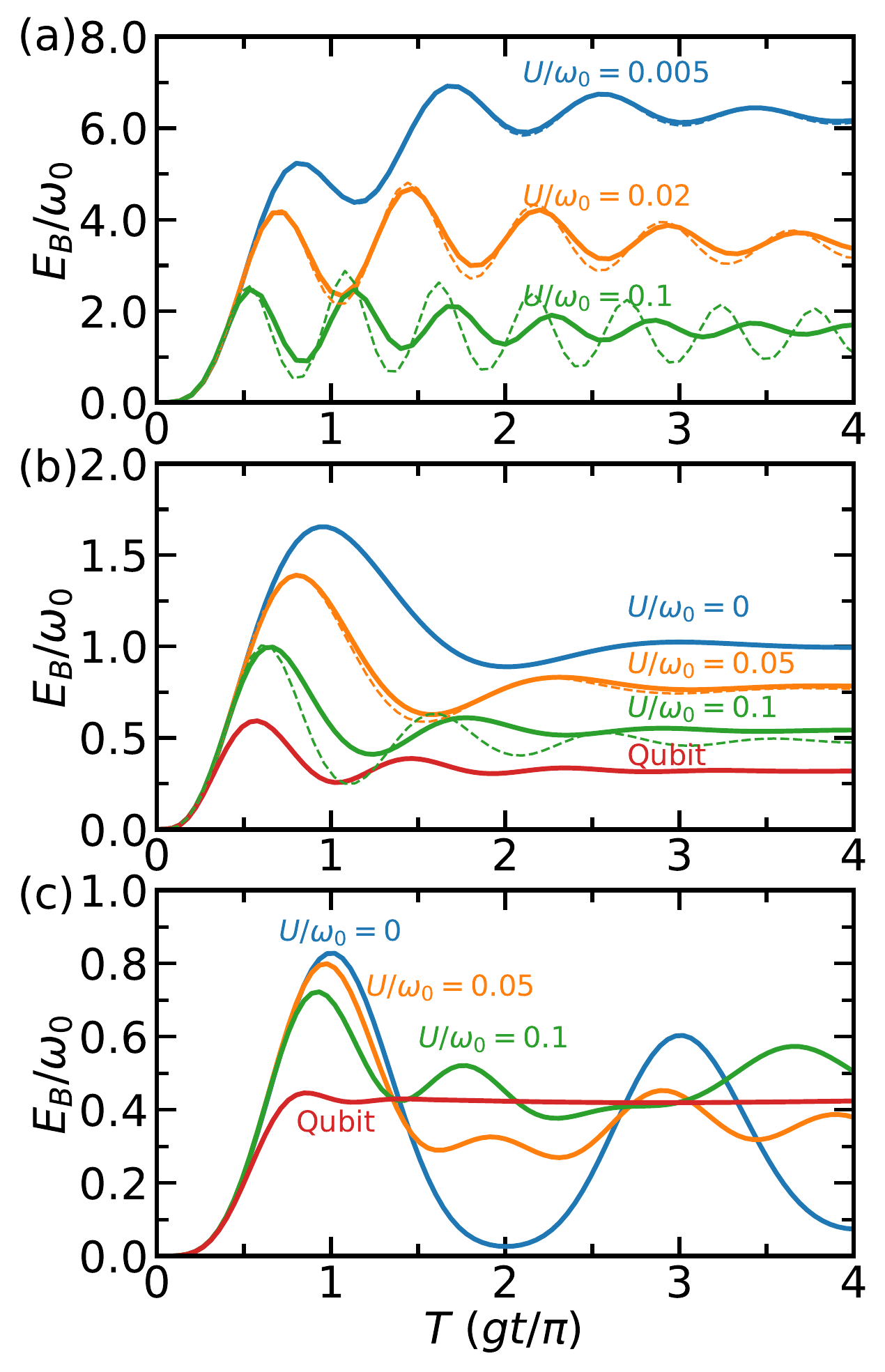}
\caption{Benchmarking the averaged stored energy $E_B/\omega_0$ as a function of charging time $T\equiv gt/\pi$ for several $U/\omega_0$. (a) The same parameters as in Fig.~\ref{fig:ew} are used. (b) The same parameters as in Fig.~\ref{fig:ew} are used, but $\gamma=\omega_0$.  The dashed lines correspond to the mean-field calculations. (c) The case of $\Delta=0$, $\gamma=0.05\omega_0$, $g=0.2\omega_0$, and $F=0.1\omega_0$.  }\label{fig:ewv}
\end{center}
\end{figure}

After charging, the battery is disconnected from the charger. We want the battery to store maximum energy within minimal charging time. Thus, the appropriate quantities to characterize the performance of the battery are the maximum stored energy and the time it takes to reach this maximum energy is referred to as optimal charging time~\cite{Andolina2018,delmonte2021characterization,crescente2020ultrafast}. The average energy stored by the battery is expressed as follows:
\begin{align}
    E_B(t) = \omega_0 \langle b^\dag b\rangle(t) + U\langle b^\dag b^\dag bb\rangle(t).\label{eq:eb}
\end{align}
The optimal charging time ($t_\textrm{max}$) is defined as the time to reach the maximum $E_B$, i.e., $E_B(t_\textrm{max})= E_\textrm{max}$~\cite{Andolina2018,delmonte2021characterization,crescente2020ultrafast}.  It should be noted that not all of the transferred energy can be used for work. The maximal amount of energy that can be extracted for work is known as ergotropy~\cite{allahverdyan04-max}. The ergotropy can be obtained by considering the energy of the passive state of the battery, from which no work can be extracted. The ergotropy of the quantum battery system can mathematically be expressed as follows~\cite{Farina2019,farina2021dissipative,tabesh2020environment,francica2020quantum,ccakmak2020ergotropy,barra2019dissipative}:
\begin{align}
    \mathcal{W}(t)=E_B(t) - \mathrm{Tr}[\rho^p(t)(\omega_0 b^\dag b + U b^\dag b^\dag bb)],\label{eq:erg}
\end{align}
where $\rho^p(t)$ is the passive state of the battery. The passive state of the battery is obtained by diagonalizing $\rho_B(t)$, which is the reduced density operator for the battery $\rho_B(t)=\textrm{Tr}_A[\rho(t)]$, and then ordering the eigenvalues such that the largest eigenvalue of $\rho_B(t)$ occupies the lowest energy level of the battery. $\rho^p$ is defined as $\rho^p\equiv\sum_n r_n|\varepsilon_n\rangle\langle\varepsilon_n|$, where $r_n>r_{n+1}$ is the eigenvalue of $\rho_B(t)$ and $\varepsilon_n<\varepsilon_{n+1}$ is the energy level of the battery~\cite{allahverdyan04-max, Farina2019} [Fig. \ref{fig:bat}(b)]. 

We solve the master equation for $\rho(t)$ and calculate $E_B$ and $\mathcal{W}$ numerically using the QuTiP package~\cite{johansson2012qutip}. In all the calculations, we truncate the sizes of our Fock space by limiting the bosonic particles associated with the quantum oscillators to $25$, since the convergence is already achieved for the parameters that we use.  In the specific case of simulating the qubit battery, we use the Hamiltonian of Eq.~\eqref{eq:H} without the Kerr nonlinearity term, but the Fock space size of the battery is taken to be two that corresponds to the two-level system.  In Figs.~\ref{fig:ew}(a) and~\ref{fig:ew}(b), we show $E_B$ as a function of time and $E_\textrm{max}$ for the qubit battery, the harmonic oscillator battery, and the Kerr battery with several $U$ values, while in Fig.~\ref{fig:ew}(c), we show $t_\textrm{max}$. The parameters used for the calculation are given in the figure caption. It is clear that the qubit battery has the lowest $E_\textrm{max}$ due to having only two levels for storing energy. However, the qubit battery reaches  $E_\textrm{max}$ faster than the other batteries. On the other hand, the harmonic oscillator battery gives the largest $E_\textrm{max}$ albeit having the longest $t_\textrm{max}$, as shown in Figs.~\ref{fig:ew}(a) and~\ref{fig:ew}(c).  Adding nonlinearity to the battery decreases the stored energy, but the optimal charging time also decreases, as shown in Figs.~\ref{fig:ew}(b) and~\ref{fig:ew}(c). Even for small nonlinearity $U=0.005\omega_0$, $t_\textrm{max}$ for the Kerr battery is half of that for the harmonic oscillator battery while keeping much higher $E_\textrm{max}$ than the qubit battery. 

To complement the QuTiP results, we solve for $E_B$ using the mean-field approximation~\cite{Zhang2021,zhang2021driven}, in which $\langle b^\dag b\rangle\approx \langle b^\dag \rangle \langle b\rangle$ and $\langle b^\dag b^\dag b b\rangle\approx \langle b^\dag \rangle^2 \langle b\rangle^2$. This approximation is accurate when the number of excited particles in the battery is large enough~\cite{zhang2021driven}. By using the master equation of Eq.~\eqref{eq:me}, we arrive at the following differential equations:
\begin{subequations}
\begin{align}
    \partial_t\langle a\rangle&=-\left(i\Delta +\frac{\gamma}{2}\right)\langle a\rangle-ig\langle b\rangle -iF,\\
    \partial_t\langle b\rangle&=-i\Delta \langle b\rangle-ig\langle a\rangle-i 2U\langle b \rangle^* \langle b\rangle^2,\label{eq:sc}
\end{align}
\end{subequations}
from which we calculate $E_B$. The mean-field solutions are given as dashed lines in Figs.~\ref{fig:ewv}(a)--(c). We found that increasing $U$ reduces the accuracy of the mean-field approximation. Physically, this reduction implies that the number of excited particles is reduced since the energy levels become more separated. Nevertheless, indicated by their similar behavior, the mean-field solutions are already consistent with the QuTiP results.  In Fig.~\ref{fig:ewv}(b), we show the case when we increase the dissipation. It is clear that $E_B$ becomes lower and there is less Rabi oscillation before reaching the steady state.  We also arrive at a similar conclusion that the Kerr battery gives better $E_B$ than the qubit battery and smaller $t_\textrm{max}$ than the harmonic oscillator battery. However, in the case of the low driving force and dissipation, i.e., $F,\gamma<g$, we find that $t_\textrm{max}$ for all batteries are not quite different as shown in Fig.~\ref{fig:ewv}(c).  We argue that such a behavior emerges because the harmonic oscillator battery and the Kerr battery are, with the low driving force, not excited to a much higher level relative to the qubit battery. 
\begin{figure}
\centering\includegraphics[clip,width=7.5cm]{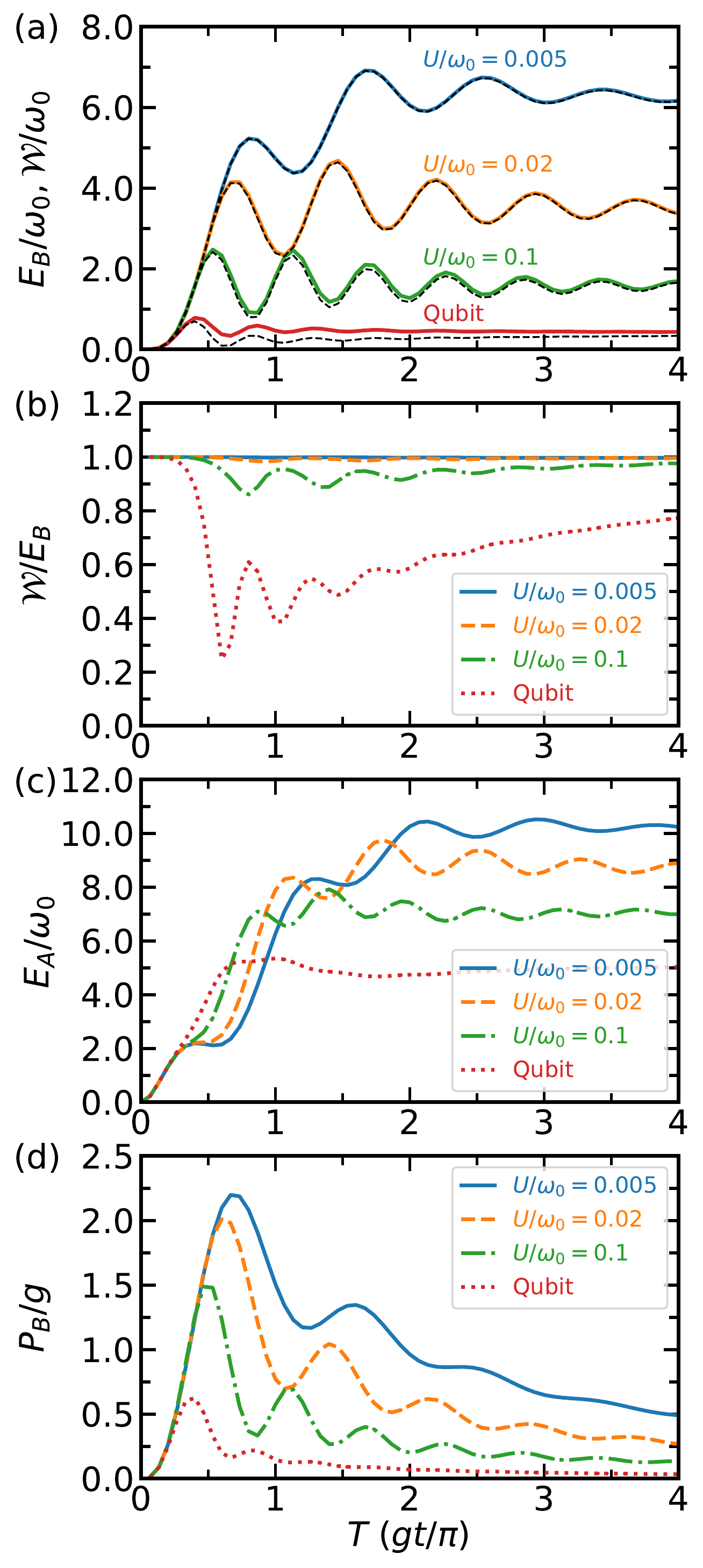}
\caption{\label{fig:pw} Performance of Kerr batteries versus qubit battery. (a) $E_B/\omega_0$ and $\mathcal{W}/\omega_0$ as a function of charging time $T\equiv gt/\pi$ for several different values of $U/\omega_0$. The corresponding $\mathcal{W}/\omega_0$ is plotted as a dashed line. (b) $\mathcal{W}/E_B$ as a function of time $T$. (c) $E_A/\omega_0$ as a function of time $T$.  (d) $P_B/g$ as a function of time $T$.}
\end{figure}

In Fig.~\ref{fig:pw}(a), we show the ergotropy $\mathcal{W}$ as a function of charging time for the qubit and Kerr batteries. It is clear that we can extract more stored energy using the Kerr battery rather than using the qubit battery. In particular, $\mathcal{W}\approx E_B$ for the Kerr battery with $U\le 0.1\omega_0$, which means that most of the stored energy can be extracted for work. The ratio between ergotropy and the stored energy is given in Fig. ~\ref{fig:pw}(b). The ratio is close to unity for the parameters we used and higher than that of the qubit battery. This suggests that the state of the system is almost a pure state~\cite{allahverdyan04-max}. In the pure state, the state of the quantum battery can be factorized into the states of charger and battery \cite{Farina2019}, and the eigenvalue of $\rho_B(t)$ is unity corresponding to vanishing passive energy [Eq.~\eqref{eq:erg}]. By increasing $U$, the energy levels of the charger and of the battery become less similar, making the state of the system not factorized. In other words, the state of the battery  becomes more mixed, which increases the passive energy. Hence, $\mathcal{W}$ gets lower than $E_B$ with increasing $U$.  

In Fig.~\ref{fig:pw}(c), we show the energy of charger $E_A$ as a function of charging time. The energy of the charger is defined as  $E_A=\omega_0 \langle a^\dag a\rangle$. Similar to $E_B$, $E_A$ oscillates before reaching a steady state.  Compared with $E_B$ in Fig.~\ref{fig:pw}(a), $E_A$ is lower than $E_B$ in the steady state due to the anharmonicity in the battery, which makes the energy levels of the battery and of the charger less similar.  Nevertheless, we can see that $E_A$ and $E_B$ are out-of-phase each other highlighting the energy transfer between the charger and the battery.  In Fig.~\ref{fig:pw}(d), we show the average charging power of the battery as a function of time, which is defined as $P_B=E_B/t$~\cite{Farina2019,Andolina2018,crescente2020ultrafast,delmonte2021characterization}. Similar to $E_B$, $P_B$ of the Kerr battery decreases with increasing $U$, but reaches maximum faster, and the qubit battery has the lowest $P_B$.  We can also see that the time to reach maximum $P_B$ is not always the same with $t_\textrm{max}$, but they become closer with increasing $U$. 

It is noted that anharmonicity of energy levels can also be realized in a superconducting qubit, the so-called transmon, in which the energy level separation decreases with increasing levels\cite{koch2007,peterer2015coherence,wang2022,place2021}.  This system corresponds to the slightly negative value of $U$ in the case of Kerr nonlinearity.  For the same magnitude of $U$, the negative $U$ might give larger stored energy, since more energy levels can be populated than in the case of positive $U$ due to the decreasing level separation. Further explorations in this direction can be of great interest in the near future.

% \section{Conclusions}
In conclusion, we show that the Kerr battery has better performance than the qubit battery in terms of the amount of stored energy, charging power, and ergotropy. The Kerr battery also has a smaller optimal charging time than the harmonic oscillator battery.  The properties of the battery system are transformed from that of the harmonic oscillator battery to a qubit battery by increasing nonlinearity.  The Kerr quantum battery in this sense takes an advantage over the qubit battery and the harmonic oscillator battery by possessing intermediate nonlinearity.

% \begin{acknowledgements}
We thank Dr.~Charles~Downing (University of Exeter) for the fruitful discussion and his suggestion about Kerr nonlinearity.  Universitas Indonesia supports this research through PUTI Q1 Research Grant No.~NKB-478/UN2.RST/HKP.05.00/2022.  The BRIN authors acknowledge Mahameru BRIN for its HPC facility.  We also thank Dr.~Juzar Thingna (University of Massachusetts Lowell) for his willingness to reply to our email correspondence when we first learned about the quantum battery.  
% \end{acknowledgements}

% \section{Results and discussion}
% \label{sec:res}

%\bibliography{references} % Produces the bibliography via BibTeX.

%aipnum4-2.bst 2019-01-14 (MD) hand-edited version of apsrev4-1.bst

\end{document}